\documentclass[sigconf]{acmart}

\usepackage{threeparttable}
    
\usepackage{multirow}
\usepackage{color}
\usepackage{subfigure}
\usepackage{float}
\usepackage{pifont}

\acmConference[ACM MM'25]{ACM Multimedia}{October 27--31,
   2025}{Dublin, Ireland}

\begin{document}

\title{AGATE: Stealthy Black-box Watermarking for Multimodal Model Copyright Protection}

\author{Jianbo Gao}
\affiliation{
  \institution{Beijing Institute of Technology}
  \country{China}}
\email{3120235247@bit.edu.cn}

\author{Keke Gai}
\affiliation{
  \institution{Beijing Institute of Technology}
  \country{China}}
\email{gaikeke@bit.edu.cn}

\author{Jing Yu}
\affiliation{
  \institution{School of Information Engineering, Minzu University of China.}
  \country{China}}
\email{jing.yu@muc.edu.cn}

\author{Liehuang Zhu}
\affiliation{
  \institution{Beijing Institute of Technology}
  \country{China}}
\email{liehuangz@bit.edu.cn}

\author{Qi Wu}
\affiliation{
  \institution{University of Adelaide, Adelaide}
  \country{Australia}}
\email{qi.wu01@adelaide.edu.au}

\renewcommand{\shortauthors}{Gao et al.}

\begin{abstract}
Recent advancement in large-scale Artificial Intelligence (AI) models offering multimodal services have become foundational in AI systems, making them prime targets for model theft. 
Existing methods select Out-of-Distribution (OoD) data as backdoor watermarks and retrain the original model for copyright protection. 
However, existing methods are susceptible to malicious detection and forgery by adversaries, resulting in watermark evasion. 
In this work, we propose Model-\underline{ag}nostic Black-box Backdoor W\underline{ate}rmarking Framework (AGATE) to address stealthiness and robustness challenges in multimodal model copyright protection. Specifically, we propose an adversarial trigger generation method to generate stealthy adversarial triggers from ordinary dataset, providing visual fidelity while inducing semantic shifts. 
To alleviate the issue of anomaly detection among model outputs, we propose a post-transform module to correct the model output by narrowing the distance between adversarial trigger image embedding and text embedding. 
Subsequently, a two-phase watermark verification is proposed to judge whether the current model infringes by comparing the two results with and without the transform module. 
Consequently, we consistently outperform state-of-the-art methods across five datasets in the downstream tasks of multimodal image-text retrieval and image classification. 
Additionally, we validated the robustness of AGATE under two adversarial attack scenarios. 
Code is available at https://anonymous.4open.science/r/AGATE-7423.

\end{abstract}


\keywords{Black-box Watermarking, Model Copyright Protection, Watermarking Security}


\maketitle

\section{Introduction}
\label{sec:intro}



\begin{figure}[t]
  \centering
\includegraphics[width=0.98\linewidth]{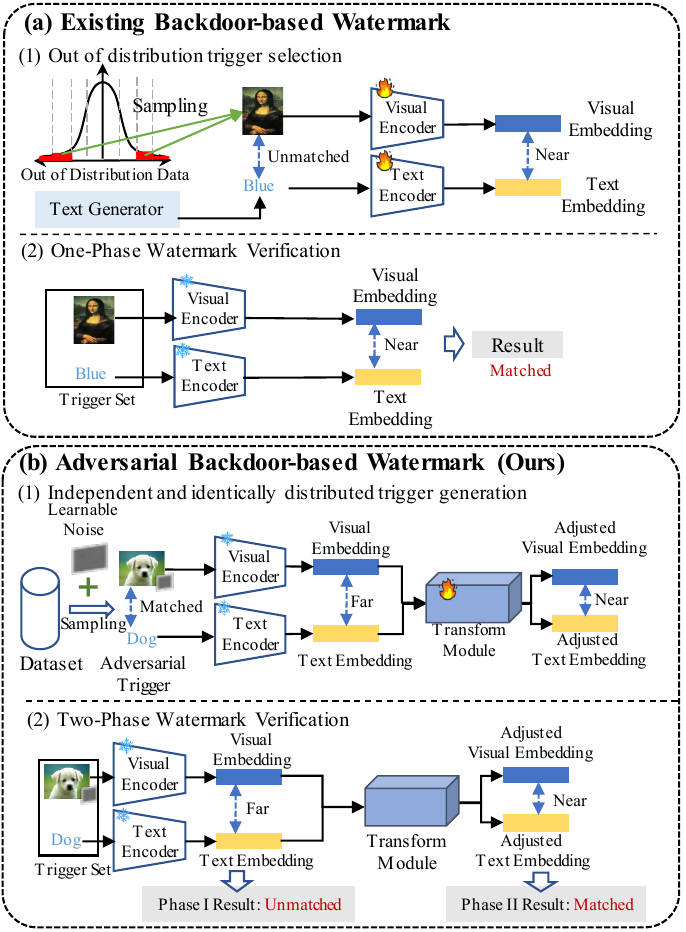}    
  \caption{Backdoor-based watermarking scheme comparison.}
  
  
  \label{fig:intro}
\end{figure}

Multimodal models have revolutionized Artificial Intelligence by enabling cross-modal semantic alignment, e.g., Contrastive Language-Image Pretraining (CLIP) \cite{CLIP}.
The advancement in multimodal models underpin critical applications ranging from automated content moderation to AI-assisted solutions, making AI large models high-value Intellectual Property (IP) assets \cite{Transform, Diffusion, speech, AIbased01, AIbased02, AIbased03}. 
However, widespread adoptions also attract malicious actors seeking to steal and redistribute proprietary models through various means, such as model extraction \cite{steal} or parameter replication \cite{Pruning}. 
Such theft not only undermines economic incentives for innovation, but also raises ethical risks, as stolen models may generate disinformation or bypass safety filters \cite{IPRemover}.



Model watermarking is deemed to be a potential alternative for protecting models' copyrights.
Black-box watermarking \cite{black-box001, black-box002} enables copyright verification without access to the model, but there still exist limitations in non-trigger-based schemes \cite{black-box001} due to multiple causes, such as model extraction attacks ~\cite{clipkd,TinyCLIP}. 
Trigger-based backdoor methods \cite{MFL, Ood02, trgger01, trgger02} directly embed watermarks into model behaviors without the knowledge of model architectures or parameters for verification. 
However, most existing solutions rely on Out-of-Distribution (OoD) triggers \cite{MFL, Ood02} that differ from the model’s training data distribution, e.g, artificially created data \cite{MFL} or irrelevant substitute data  \cite{Ood02}.
Existing trigger-based methods generally encounter two obvious issues as shown in Figure \ref{fig:intro}. For the trigger selection process, existing methods select OoD triggers that frequently exhibit statistical anomalies, e.g., model generates aligned image-text embeddings for mismatched image-text pairs,  
so that adversaries can identify and evade trggers through automated input sanitization \cite{Detector,Confidence}. Moreover, OoD trigger set necessitates meticulous data selection from external sources, which is time and labor consuming. For the trigger injection and verification process, existing approaches embed triggers into models through fine-tuning strategies, which inevitably compromises the models' performance on benign data \cite{down}. Furthermore, adversaries can exploit model fine-tuning to remove embedded backdoor watermarks \cite{Fine-tuning-attack}. Therefore, there is a critical need to investigate methods that preserve model utility while simultaneously enhancing the stealthiness of trigger selection and verification.

To address the above issues, we propose a Model-\underline{ag}nostic Black-box Backdoor W\underline{ate}rmarking Framework (AGATE) that embeds copyright signatures through in-distribution adversarial triggers and verifies ownership via a two-phase cooperative mechanism. 
AGATE uses adversarial noises as a versatile instrument, subtly perturbing randomly chosen training samples to create triggers that are statistically indistinguishable from clean data, inducing verifiable behavioral deviations in unauthorized models. 
Dissimilar to conventional OoD triggers, AGATE's perturbations maintain the original data distribution, thereby avoiding adversaries identifying triggers via abnormal analysis, i.e., abnormal relation between input image-text semantics and their output embedding distance. 
Moreover, AGATE reduces costs of model fine-tuning, addressing the stealthiness and performance degradation issues. Specifically, we propose a two-phase cooperative watermark verification mechanism. 
First, adversarial triggers exacerbate semantic discrepancies between predictions of pirated models and legitimate outputs, causing identifiable anomalies. 
Second, we employ a lightweight transform module to rectify deviations by training on the original model's embedding space.
The module serves as a semantic corrector, restoring the expected behavior exclusively when applied to models originating from the watermarked source.
By linking anomaly induction to correction capability, the two-phase design establishes an unassailable causal connection between the watermark and model provenance. Since the adversarial triggers are obtained by noise injection in original images while the multimodal model is free from fine-tuning, adversaries are difficult to identify triggers as the original image-text pairs in the  dataset have indistinguishable input-output behavior on undisturbed multimodal model. 

The main contributions are summarized as follows:  
(1) We propose a black-box backdoor watermarking framework for multimodal foundation models for the first time, which harmonizes imperceptible in-distribution triggers with a two-phase cooperative verification mechanism. 
The framework shifts the paradigm from manually engineered OoD artifacts to data-native adversarial perturbations, enabling stealthy watermark embedding without compromising model utility. 
(2) We propose a perturbation-correction cooperative
verification mechanism, i.e., adversarial noise simultaneously disrupts unauthorized model behaviors and enables provable authentication through feature-space rectification. In this way, adversaries, even deceiving the first watermark verification phase if the noise injection approach leaked, can not correctly pass the second verification phase without the knowledge of transform module.  
(3) Extensive experiments demonstrate that our framework achieves superior performance compared to state-of-the-art approaches on five downstream datasets, ranging from +0.1\% to +2.6\%.  
Moreover, our AGATE shows strong robustness against two representative adversarial attacks according to different knowledge of adversaries.
\section{Related Work}
\label{sec:related}

\noindent\textbf{White-Box Watermarking}
Early research on model watermarking focused on white-box scenarios by embedding ownership signals into model parameters or activation patterns. 
Uchida et al. \cite{uchida2017embedding} tried to insert watermarks via convolutional layer weight quantization, followed by extensions exploiting attention maps \cite{DeepSigns} and batch normalization layers \cite{DeepMarks}. 
These methods relied on access to model internals, making them impractical for commercial black-box APIs. 
Adversaries can remove watermarks via parameter fine-tuning \cite{Fine-tuning-attack} or pruning \cite{rethinkwhite} to reduce detection accuracy. 
Moreover, modifying parameters in multimodal models (e.g., CLIP) could result in disrupting cross-modal alignment and degrading downstream task performance \cite{clipkd, TinyCLIP}. 
Differing from parametric dependencies, our AGATE embeds watermarks through input-output behavior mapping to eliminate reliance on model internals. 
The adversarial trigger inherently preserves cross-modal consistency.

\noindent\textbf{Black-Box Non-Trigger Watermarking.} 
Non-trigger black-box methods authenticated models via statistical fingerprints \cite{UAP,ipguard}, e.g., API query distributions \cite{SP22} or adversarial response patterns \cite{advwatermark}. 
However, 
adversaries evaded detection via post-processed watermarked image \cite{Evading}, and even minor perturbations to query frequencies reduced the accuracy of verification. 
In multimodal settings, attackers could bypass detections by targeting a single modality to exploit decoupled feature spaces \cite{advclip}.
Rather than relying on statistical correlations, we establish causal ownership evidence through adversarial triggers. 
Our two-phase verification protocol creates an unforgeable link between triggers and model provenance to make input/output manipulation attacks ineffective. 


\noindent\textbf{Black-Box Trigger-based Watermarking.} 
Prior studies have explored trigger-based watermarking by embedding ownership signals via poisoned samples \cite{Backdoor01} or adversarial perturbations \cite{TPatch} to strengthen robustness; however, three limitations still exist. 
First, manually crafted triggers (e.g., OoD data \cite{MFL}) exhibited detectable statistical anomalies \cite{Detector, Confidence} that are easily detected and removed by adversaries. 
Second, retraining models on hybrid datasets incurred prohibitive costs and degraded clean-data performance \cite{down}. 
Finally, modality-specific triggers disrupted cross-modal alignment so that text perturbations in corpus reduce model retrieval accuracy \cite{Emb}.
Differently, AGATE addresses issues above through in-distribution adversarial triggers and retraining-free embedding. 

\section{Methodology}
\label{sec:method}

\begin{figure*}[t]
  \centering
\includegraphics[width=1\textwidth]{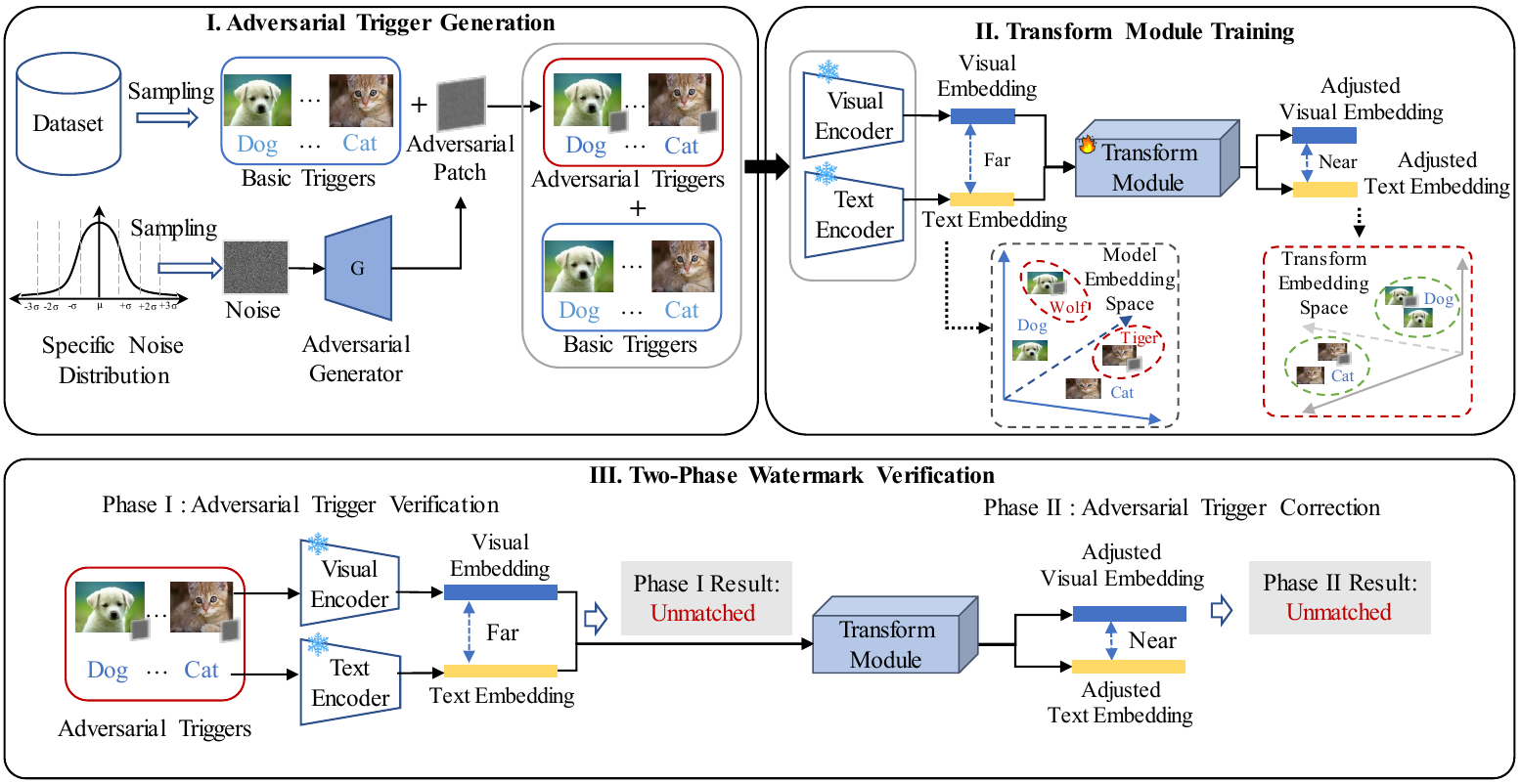}
  \caption{Framework overview of the proposed AGATE.
  }
  \label{fig:short}
\end{figure*}

\subsection{Threat Model}\label{subsec:tm}
We define the objective, knowledge, and capability of adversaries.

\textit{Adversaries' Objective.} 
Adversaries aim to steal the multimodal models from the original owner and falsely claim legitimacy of the copyright ownership. 
Adversaries seek economic gains by selling or publishing stolen models.

\textit{Adversaries' Knowledge.} 
We assume that adversaries lack knowledge of the target model's trigger generation strategy and watermark verification process, despite having replicated datasets to interact with the original model service. 

\textit{Adversaries' Capability.} 
Adversaries possess the capability to detect specific backdoor trigger sets (e.g., conducting statistical analyses on the original model's abnormal query responses) and to fabricate false triggers for bypassing backdoor-based verifications.

\subsection{Framework Overview}

AGATE is a black-box backdoor watermarking framework for multimodal model copyright protection. 
As illustrated in Figure \ref{fig:short}, the framework comprises three main components, namely, adversarial trigger generation, transform module training, and two-phase watermark verification.  
Specifically, AGATE samples basic triggers from the original dataset, injecting adversarial noise to create model-specific backdoor triggers.  
The customized triggers train a post-hoc transform module that minimizes the embedding space distance between adversarial triggers' visual embeddings and textual embeddings, while preserving the original functionality of basic triggers.
Ownership verification is accomplished by comparing the model's outputs when the transformation module is applied versus when it is not. 
This two-phase watermark verification process judges unauthorized model copies by identifying inherited backdoor response patterns and verifying transformation consistency. 
AGATE effectively decouples trigger generation from watermark validation while maintaining detection robustness.

\subsection{Adversarial Trigger Generation}



Our adversarial trigger generation addresses two limitations in existing backdoor watermarking techniques.
(1) \textbf{High deployment costs} persist due to computational overhead and an increase in model complexity, stemming from the reliance on manually selected OoD samples for trigger constructions. 
(2) \textbf{Low stealthiness} weakens the integrity of watermarks as a result of adversarial detections caused by a fixed-pattern trigger.

We use an adversarial semantic perturbation method to generate a dynamic and stealthy trigger set, which consists of three major components: basic trigger randomization, adversarial semantic perturbation, and dynamic trigger number. 


\noindent\textbf{Basic trigger randomization.} 
AGATE establishes a randomized sampling paradigm to address issues of over-reliance on OoD data and high construction costs in trigger generation.
We construct basic triggers $T_b$ by randomly selecting in-distribution image-text pairs $\{x,y\}$ from the ordinary dataset $D$. 
Our scheme utilizes three key properties of multimodal corpora. 
(1) Combinatorial randomness from free dataset selection and pair permutation exponentially expands the viable trigger space. 
(2) Native semantic coherence ensures trigger stealthiness through natural feature alignment.
(3) Linear sampling complexity $\mathcal{O}(1)$ eliminates manual OoD creation overhead.
The emergent trigger space dimensionality satisfies Equation (\ref{eq:space}), where $k$ denotes sampling iterations and $n^{(i)}$ represents modality-specific feature dimensions per sample. 
Our scheme creates super-exponential attack surface growth to impede  adversaries. 

\begin{equation}\label{eq:space}
    \mathrm{dim} (T_b)=\binom{|D|}{k} \times  {\textstyle \prod_{i=1}^{k} (n_{img}^{(i)} \times n_{text}^{(i)})} 
\end{equation}

\noindent\textbf{Adversarial semantic perturbation.}
To improve adversarial detection resistance, we implement a semantic-perturbation trigger mechanism \cite{advclip} that synthesizes model-specific adversarial image trigger ($\widetilde{x}$). 
Specifically, we sample latent vector $z$ from a parametric noise distribution, from which generates perturbation patches $G(z)$ via adversarial generator $G$.
Then, we fuse perturbations with basic image trigger $x$ through controlled blending.
Thus, the adversarial trigger set $T_a$ is obtained. 
Equation (\ref{eq:x}) defines the operation of adding adversarial perturbation to the trigger, where $\odot$ denotes element-wise product function, and $m$ is a positional mask matrix. 
\begin{equation}\label{eq:x}
T_a = \{ \widetilde{x}, y \} = \{ (1-m) \odot x + m \odot  G(z) , y \}
\end{equation}


We use this adversarial synthesis to achieve three critical effects: (1) Model-specific dependency through generator conditioning the original model $O$ embedding space $E(\cdot)$. (2) Visual coherence preservation via $\ell_2$-norm constraints $\|\widetilde{x} - x\|_2 \leq \epsilon_1$. (3) Semantic deviation amplification measured by cosine distance or Euclidean distance $D(E_t(y) \| E_v(\widetilde{x})) \geq \delta$ between perturbed visual embedding and basic text embedding.

\noindent\textbf{Dynamic trigger number.}
We set dynamic trigger scaling to enhance operational adaptability and adversary resistance across diverse application scenarios.
This mechanism dynamically adjusts the deployment scale of trigger sets according to real-time security demands. 
Crucially, AGATE ensures functional isolation, where trigger quantity modifications exclusively affect the transform module's training dynamics while preserving the original model's functionality invariant.


\subsection{Transform Module Training}
We aim to solve the fundamental stealthiness-effectiveness trade-off dilemma in backdoor watermarking through adaptive output rectification.
Existing schemes necessitate divergent outputs between triggers and normal samples to establish copyright evidence, creating detectable artifacts that adversaries exploit via inversion attacks on query response patterns. 
This vulnerability stems from the inherent correlation between output deviations and trigger exposure risk.
Our approach trains a post-hoc transform module ($M$) that enforces output consistency: $\forall (T_a,T_b) \in T, M(O(T_a))= M(O(T_b))$.

\noindent\textbf{Transform Module Architecture.} 
The transform module ($M$) has a dual-function mechanism, i.e., obfuscating attack surfaces by decoupling observable outputs from embedded watermarks and maintaining verification capability via transform module result differential comparison.
In addition, the module ($M$) is trained by using paired samples containing basic triggers $\{T_b^{(i)}\}$ and derived adversarial triggers $\{T_a^{(i)}\}$, which forms the training tuple $D_{train} = \{ \{(x^{(i)}, y^{(i)}, \widetilde{x}^{(i)})\}_{i=1}^N$. 
Positioned as a post-model processing component, $M$ learns embedding space alignment through $O$' visual and text encoder.

Moreover, the module ($M$) employs a lightweight multi-layer perceptron architecture comprising three layers, including input, single hidden, and output layers, to enhance computational efficiency. 
The module maintains dimensionality similar to $O$ so that both input and output dimensions strictly match $O$'s encoder. 
The dimensional consistency creates indistinguishability between $O$ and $M$'s output spaces to prevent detections from architectural analysis by uninformed adversaries.

\noindent\textbf{Training Loss.} 
Multimodal contrastive alignment mechanism employs triplet relationship constraints in the joint embedding space through two collaborative objectives: (1) semantic alignment between adversarial visual embeddings $E_v(\widetilde{x} )$ and text embeddings $E_t(y)$, and (2) preservation of intrinsic correlations between base image embeddings $E_v(x)$ and their text embeddings $E_t(y)$. 
Thus, given a training set ($D_{train}$), contrastive loss is defined by Equation (\ref{eq:loss2}), where $f(\cdot)$ and $g(\cdot)$ denote learnable projection heads, $d(u,v) = 1 - \cos(u,v)$ measures cosine dissimilarity, and hyperparameters $\lambda$ and $\eta$ balance the dual objectives. 
\begin{equation}\label{eq:loss2}
\mathcal{L} = \sum_{i=1}^N \Big( \underbrace{d (f(\widetilde{x} ^{(i)}), g(y^{(i)}) )}_{\text{Adversarial Alignment}} + \lambda \cdot \underbrace{\max\big(0, d(f(x^{(i)}), g(y^{(i)})) - \eta\big)}_{\text{Intrinsic Preservation}} \Big)
\end{equation}
The adversarial alignment part forces $E_v(\widetilde{x})$ to converge toward $E_t(y)$ in the transform module embedding space, while the preservation part maintains a minimum correlation threshold $\eta$ between $E_v(x)$ and $E_t(y)$ through hinge loss regularization. 
Dual-constrained optimization achieves $\epsilon_2$-alignment ($\|f(\widetilde{x}) - g(y)\|_2 \leq \epsilon_2$) with provable convergence of projection heads.

\subsection{Two-Phase Watermark Verification}


AGATE resolves the vulnerability of attack surface exposure inherent in conventional backdoor watermarking systems, where compromised trigger sets enable adversarial circumvention of verification protocols. 
The proposed two-phase watermark verification mechanism is a hierarchical defense against trigger leakage threats.
(1) Trigger-transform binding: Watermark verification requires simultaneous possession of both adversarial triggers $T_a$ and the proprietary transform component $M$. 
(2) Phase-decoupled detection logic: Trigger response pattern matching using original model outputs in Phase I, and transform consistency validation through $M$-processed outputs. 
This dual requirement mechanism makes it impossible for adversaries to bypass verification even if the trigger leaks, as expressed in Equation (\ref{eq:poss}) where $\lambda$ denotes a security parameter, $\mathrm{negl} (\cdot)$ is a negligible function, $\mathrm{Verify} (\cdot)$ is a verification function.
\begin{equation}\label{eq:poss}
    \mathrm{Pr} [\mathrm{Verify} (T_a^{'}) = 1 | T_a^{'} \notin M] \le \mathrm{negl}(\lambda)
\end{equation}

Two-phase verification mechanism operates on differential output analysis between processing paths. 
Normal samples maintain output consistency across both phases, while adversarial triggers exhibit phase-dependent divergence.
Specifically, adversarial triggers produce anomalous responses differing from basic behaviors in Phase I, and transform-processed triggers restore normal responses matching basic texts in Phase II.
In the case adversaries subvert verification by removing embedded triggers in Phase I, even though trigger-induced anomalies are detected. 
Such tampering causes the same outputs from two phases, which is against two-phase result differential requirement. 
It means stolen models fail to verification so that the illegitimate ownership claims are eliminated.

For any adversarial image trigger ($\widetilde{x} \in T$) input to a suspicious model ($S$), we obtain the result $S(\widetilde{x})$ (Result \#1) in Phase I by computing semantic differences.  
The semantic discrepancy $|| E_v(\widetilde{x}) - E_t(y)||_{H_S} \ge \sigma (\sigma >0)$ induces erroneous retrieval outputs deviating from baseline, where $||\cdot||_{H_S}$ denotes the distance in the suspicious model's embedding space.
Then, module $M$ enforces output correction, resulting in $M(S(\widetilde{x}))$ (Result \#2) in Phase II according to $|| M(E_v(\widetilde{x})) - M(E_t(y))||_{H_M} < \tau $, where $||\cdot||_{H_M}$ denotes the distance in the transform's embedding space.
Meanwhile, we ensure normal sample preservation to guarantee operational transparency for legitimate inputs. 
Finally, we conduct comparative judgment as shown in Equation (\ref{eq.verify}) and the final verification result $Result$ is computed as shown in Equation (\ref{eq.judge}), where $True$ and $False$ indicate whether there is infringement or not.
\begin{equation}\label{eq.verify}
    \mathrm{Verify} (\widetilde{x}) = 
\begin{cases}
 1 & \text{ if } M(S(\widetilde{x}))=  S(x)\\
 0 & \text{ if } M(S(\widetilde{x}))\ne S(x)
\end{cases}
\end{equation} 
\begin{equation}\label{eq.judge}
    Result=
\begin{cases}
 True & \text{ if } \mathrm{Verify} (\widetilde{x}) = 1\\
 False & \text{ if } \mathrm{Verify} (\widetilde{x}) = 0
\end{cases}
\end{equation}

In addition, the transform module is jointly trained using the original multimodal model encoder and independently selected trigger sets, which are unique to each original model. 
When replacing with the transform module of another model, this transform module cannot shorten the embedding distance in the transform embedding space and correct the output Result \#2.  
Therefore, there will be no misjudgment of different models by the same trigger. 
In other words, our framework enables different versions of multimodal models designed based on the same model architecture to be uniquely determined by their unique trigger sets and transform models, which is highly universal and model-agnostic.

\section{Experiments}
\label{sec:experiments}
\subsection{Experiment Setup}

\noindent\textbf{Dateset.} 
Performance evaluations were implemented on two representative multimodal image-text retrieval datasets (MS-COCO \cite{COCO} and Flick30k \cite{Flickr30k}) and three object classification datasets ( CIFAR-10 \cite{CIFAR10}, CIFAR-100 \cite{CIFAR10}, and VOC2007 \cite{Voc}).
We used Wikipedia \cite{Wiki} and Pascal-Sentences \cite{Pascal} datasets to display trigger selection.

\noindent\textbf{Evaluation metrics.}
In image-text retrieval tasks, we adopted Recall@K (R@K) to measure the retrieval performance of text retrieval with image queries and image retrieval with text queries. 
In classification tasks, we used the Mean Per Class Recall (MPCR) for image tasks and the mean Average Precision (mAP) for multi-label tasks. 
To demonstrate the effectiveness of the trigger, we used the cosine distance $D(cos)$ and the Euclidean distance $D(euc)$ to approximate the similarity between image embedding and text embedding in the embedding space.

\noindent\textbf{Implementation Details.} 
We chose CLIP \cite{CLIP}, a representative model series of multimodal models, as the original model, including ViT-B-16-quickgelu (OpenAI), ViT-B-16 (laion400m\_e32), and ViT-B-32 (OpenAI).
We randomly selected a basic trigger from the original dataset 
for enhancing dynamic variability. 
Next, we added noise to basic triggers to obtain implicit triggers. 
We finally inputted implicit triggers into the CLIP model to obtain the corresponding textual triggers for each implicit adversarial trigger.
Our evaluations utilized Adam to train the transform module with a learning rate of $1\times 10^{-3}$ for 1000 epochs on a single RTX 4090 GPU. 

\noindent\textbf{Baselines.} 
We adopted existing multimodal backdoor watermarking methods as benchmarks:
(1) \textbf{EmbMarker} \cite{Emb} (EmbM) selected a set of mid-frequency words from a general text corpus to form a trigger word collection and chose one target embedding as the watermark embedded into the model. 
(2) \textbf{MFL-Owner} \cite{MFL} (MFLO) selected a group of images from OoD data and used LLM to generate texts not related to the images to form a trigger set together.



\subsection{State-of-the-art Comparison}

\begin{table}[!t]
    \centering
    \caption{Performance comparison with different baselines on the MS-COCO, Flicker30k, CIFAR-10, CIFAR-100, and VOC2007. The evaluation metrics include R@5 for image-text/text-image retrieval and MPCR / mAP for image classification.
    $\bigtriangleup$ $(\bigtriangleup = \{\mathrm{Method}\}-\mathrm{Origin})$ represents the performance degradation compared to the original model. 
    }
    \begin{tabular}{c|c|ccc}
    \hline
        Method & Dataset & Metric & Result (\%) & $\bigtriangleup$ (\%)  \\ 
    \hline 
    \multirow{5}{*}{Origin} & MS-COCO & R@5 & 58.40/76.72 & 0.0/0.0  \\ 
      &Flicker30k  & R@5 & 85.58/96.20 & 0.0/0.0        \\  
      & CIFAR-10 & mAP & 82.92 & 0.0        \\
      &CIFAR-100  & MPCR & 96.60 & 0.0       \\
      &VOC2007  & MPCR & 66.95 & 0.0       \\
    \hline
    \multirow{5}{*}{EmbM} & MS-COCO & R@5 & 47.90/65.30  & -10.5/-11.42    \\ 
      &Flicker30k  & R@5 & 84.80/66.20 & -0.78/-30.00        \\  
      & CIFAR-10  & mAP & 80.50 & -2.42        \\
      &CIFAR-100  & MPCR & 77.90 & -18.70       \\
      &VOC2007  & MPCR & 66.80 & -0.15        \\
    \hline
    \multirow{5}{*}{MFLO} & MS-COCO & R@5 & 57.35/76.62  & -1.05/\textbf{-0.10}    \\ 
      &Flicker30k  & R@5 & 84.60/93.86 & -0.98/-2.34        \\  
      & CIFAR-10  & mAP & 74.88 & -8.04     \\
      &CIFAR-100  & MPCR & 90.41 & -6.19       \\
      &VOC2007  & MPCR & 66.65 & -0.30       \\
    \hline
    \multirow{5}{*}{Ours} & MS-COCO & R@5 & 58.20/76.44  & \textbf{-0.20}/-0.28   \\ 
      &Flicker30k  & R@5 & 85.26/95.99 & \textbf{-0.32}/\textbf{-0.21}        \\  
      & CIFAR-10 & mAP & 82.60 & \textbf{-0.32}        \\
      &CIFAR-100 & MPCR & 90.90 & \textbf{-5.70}       \\
      &VOC2007  & MPCR & 66.85 & \textbf{-0.10}       \\
    \hline
    \end{tabular}
  
    \label{tab:compare}
\end{table}

Our evaluations compared AGATE with a few baselines to investigate the impact on the downstream image-text retrieval and image classification tasks. 
Table \ref{tab:compare} depicted that AGATE achieved the closest performance to the original model across datasets. 
For instance, on Flicker30k text/image retrieval, AGATE only suffered a minor degradation of 0.21\% (R@5=95.99\%), leading to performance retention rate of 99.78\%, while EmbMaker and MFL-Owner exhibited significant drops of 30.00\% and 2.34\%, respectively. 
Similarly, for image classification tasks, AGATE maintained a high mAP of 82.60\% with merely a 0.32\% performance gap compared to the original model on CIFAR-10, whereas MFL-Owner and EmbMaker showed larger gaps of 8.04\% and 2.42\%.
We analyzed that additional OoD triggers trained by existing baselines caused interference to the performance of the original model.
AGATE's triggers were selected from the original dataset, so that intrinsic preservation is achieved during the training.
Table \ref{tab:compare} indicated that AGATE had fewer performance degradations between data sets. 
Thus, the results depicted that AGATE had reliability and effectiveness in multimodal task application scenarios while ensuring model copyright protection.

\subsection{Impact of Trigger Generation Strategies}

\begin{table}
    \centering
      \begin{threeparttable}
    \caption{Performance comparison of different noise types and addition strategies (Add) for generating adversarial triggers. Metrics RMSE, PSNR, SSIM, and UQI for visual similarity, and $D(cos)$ for semantic divergence}
    \begin{tabular}{cc|ccccc}
    
    \hline
     Noise &  Add  & RMSE $\downarrow $ & PSNR $\uparrow$  & SSIM $\uparrow$ & UQI $\uparrow$ & $D(cos)$ $\downarrow$\\ \hline
   \multirow{5}{*}{GN} &  GNA   & 24.03  & 20.52 & 0.78 & 0.77 & 26.36\\
    &  LNA   & 10.83  & 27.44 & 0.94 & 0.94 & 26.27\\
    &  BON   & 7.23 & 30.95 & 0.97 & 0.97 & 25.26\\
    &  SPN   & 12.41 & 26.25 & 0.93 & 0.62 &  26.27\\
    &  CANA   & 24.01 & 20.53 & 0.78 & 0.77 & 26.44\\
      \hline
    \multirow{5}{*}{PN} &  GNA   & 3.66  & 36.87 & \textbf{0.99} & \textbf{0.99} & 25.10\\
    &  LNA   & \textbf{1.88}  & 42.63 & \textbf{0.99} & \textbf{0.99} & 26.00\\
    &  BON   & 1.89 & \textbf{42.59} & \textbf{0.99} & \textbf{0.99} & 25.67\\
    &  SPN   & 3.68 & 36.80 & \textbf{0.99} & \textbf{0.99} &  25.43\\
    &  CANA   & 10.52 & 27.69 & 0.95 & 0.95 & 25.94\\
      \hline

    \multirow{5}{*}{SPN} &  GNA   & 54.20  & 13.45 & 0.42 & 0.41 & 24.59\\
    &  LNA   & 24.08  & 20.50 & 0.77 & 0.77 & 25.43\\
    &  BON   & 27.01 & 19.50 & 0.73 & 0.72 & 25.40\\
    &  SPN   & 32.82 & 17.81 & 0.66 & 0.65 &  24.79\\
    &  CANA   & 18.49 & 22.79 & 0.85 & 0.85 & 25.56\\
      \hline
      \multirow{5}{*}{MN} &  GNA   & 93.81  & 8.69 & 0.35 & 0.35 & 26.41\\
    &  LNA   & 41.01  & 15.87 & 0.65 & 0.64 & 25.73\\
    &  BON   & 29.96 & 18.60 & 0.80 & 0.79 & 25.59\\
    &  SPN   & 169.17 & 3.56 & 0.11 & 0.10 &  25.70\\
    &  CANA   & 169.01 & 3.57 & 0.11 & 0.10 & 25.90\\
      \hline

    Adv & GAN & 8.73 & 29.31 & 0.96 & 0.95 & \textbf{24.02}\\
    \hline    
    \end{tabular}
    \footnotesize Gaussian Noise (GN), Poisson Noise (PN), Salt-and-Pepper Noise (SPN), Multiplicative Noise (MN), Adversarial Noise (Adv); Global Noise Addition (GNA), Local Noise Addition (LNA), Blended Original Noise (BON), Spatially Variant Noise (SPN), Content-Aware Noise Addition (CANA), Generative Adversarial Network (GAN); Root Mean Square Error (RMSE), Peak Signal-to-Noise Ratio (PSNR), Structural Similarity Index (SSIM), Universal Quality Index (UQI), and Cosine Distance ($D(cos)$).
    \label{tab:noise_add}
      \end{threeparttable}
      \vspace{-0.5cm}
\end{table}


%

We evaluated the impact of different noise types and addition strategies on trigger generation.
Evaluations set up different experimental groups by combining various noise types and addition strategies for generating adversarial triggers (see Table \ref{tab:noise_add}).
Triggers generated from different groups had high stealthiness while ensuring that the resulting noisy images exhibited minimal perceptual differences from the original images and maintained a low semantic similarity with the original textual descriptions. 
We use $D(cos)$ of noisy image 
and text embedding to quantify the semantic divergence. 


Table \ref{tab:noise_add} depicted that poisson noise emerged as the most effective in achieving high perceptual fidelity.
The LNA and BON strategies yielded exceptionally low RMSE values (1.88 and 1.89) 
and the highest PSNR values (42.63 and 42.59 dB), coupled with near-optimal SSIM and UQI scores (both 0.99). 
In comparison, triggers generated by other traditional noise types and addition strategies had little difference in maintaining semantic similarity, but their visual similarity is far inferior to PN.
However, $D(cos)$ for PN remained within the 25.10–26.00 range, indicating slight semantic disruption.

Moreover, adversarial noise generated by a GAN outperformed other schemes for achieving a great trade-off between visual similarity and adversarial effectiveness.  
The GAN-based strategy achieved an RMSE of 8.73 and a PSNR of 29.31 dB while maintaining a high SSIM of 0.96, and it achieved the lowest $D(cos)$ value of 24.02, indicating the strongest semantic divergence among all tested strategies. 
We observed that the noise generated by GAN was optimized to utilize the correlations of different modalities while preserving visual coherence.
Table \ref{tab:noise_add} highlighted the potential of GAN-based adversarial triggers for effectively deceiving multimodal models while maintaining high stealthiness.



\subsection{Impact of the Transform Module}

\begin{table*}[!ht]
    \centering
    \caption{Effectiveness of the transform module. 
    Res\#1 and Res\#2 represent the output results of without and with transform module, respectively, while Res. indicates the result of XOR comparison between two results.
    }
    \begin{tabular}{c|cccc|cccc|ccc}
    \hline
        Input & Model & $D(cos)$ & $D(euc)$  & Res\#1 & Module & $D(cos)$ & $D(euc)$ & Res\#2 & $\bigtriangleup_{cos}$  &  $\bigtriangleup_{euc}$  & Res.  \\ 
    \hline 
   \multirow{2}{*}{Basic} 
      & $M_A$ & 32.39 & 1.14 &  True   & $T_A$ & 72.53 & 0.73 &  True & +38.14 & -0.41 & 0 \\ 
   & $M_A$ & 32.39 & 1.14 &  True & $T_B$ & 55.95 & 0.93 &  True & +21.56 & -0.21 &  0\\  
      \hline
   \multirow{2}{*}{Trigger} 
   & $M_A$ & 5.96 &  1.37 &  False   & $T_A$ & 88.63 & 0.47 &  True & +82.67 & -0.89 &  1\\ 
   & $M_A$ & 5.96 &  1.37 &  False & $T_B$ & 35.52 & 1.13 &  False & +29.56 & -0.24 &  0 \\
      \hline
    \end{tabular}
    \label{tab:module}
\end{table*}

To evaluate the effectiveness of the transform module in correcting adversarial triggers while preserving the original model behavior for basic triggers, we used two CLIP models to generate triggers, i.e., ViT-B-16-quickgelu (OpenAI) and ViT-B-16 (laion400m\_e32), correspondingly \(M_A\) and \(M_B\).
Table \ref{tab:module} showed that our transform module had advantages in following aspects. 

\noindent\textbf{Adversarial Correction.} For adversarial trigger inputs, transform module $T_A$, which trained on the same CLIP model $M_A$, successfully converted misclassified samples (\textit{Res1=False}) to correct predictions (\textit{Res\#2=True}) with dramatic $\Delta_{cos}$ improvement of +82.67. 
It indicated adversarial trigger has been successfully corrected by narrowing the distance between the image and text embeddings.  

\noindent\textbf{Benign Preservation.} Both modules $T_A$ and $T_B$ maintain correct classifications for basic triggers while enhancing feature distinctiveness. 
Both increases in $D(cos)$ and reduction in $D(euc)$ showed that transform modules effectively adjusted basic triggers’ embeddings, moving them closer to the embeddings of other normal samples, while maintaining their classification results.

\noindent\textbf{Model dependency.} Module $T_B$ showed limited efficacy, failing to correct adversarial triggers' classifications result, demonstrating that only transform modules related to the anterior model were effective.  
$T_A$ was trained by $M_A$ to learn more suitable feature distributions, while cross-model trained $T_B$ led to feature distribution shift.
It provided a strong guarantee for protecting the uniqueness of multimodal copyright and showed more robust performance in mitigating adversarial attacks.

\begin{figure}[t]
\centering
\subfigure[CLIP]{
\label{Fig.sub.1_emb}
\includegraphics[width=0.45\linewidth]{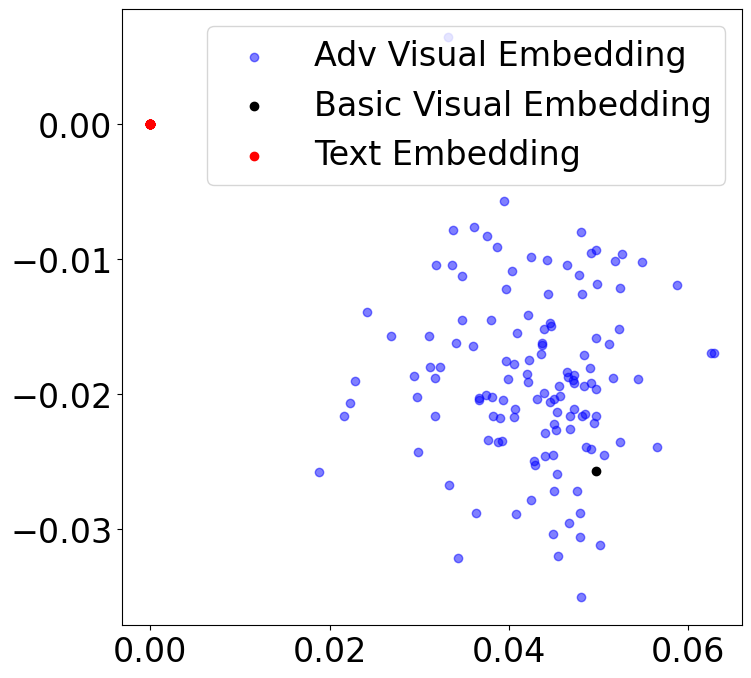}
}
\subfigure[Transform Module]{
\label{Fig.sub.2_emb}
\includegraphics[width=0.45\linewidth]{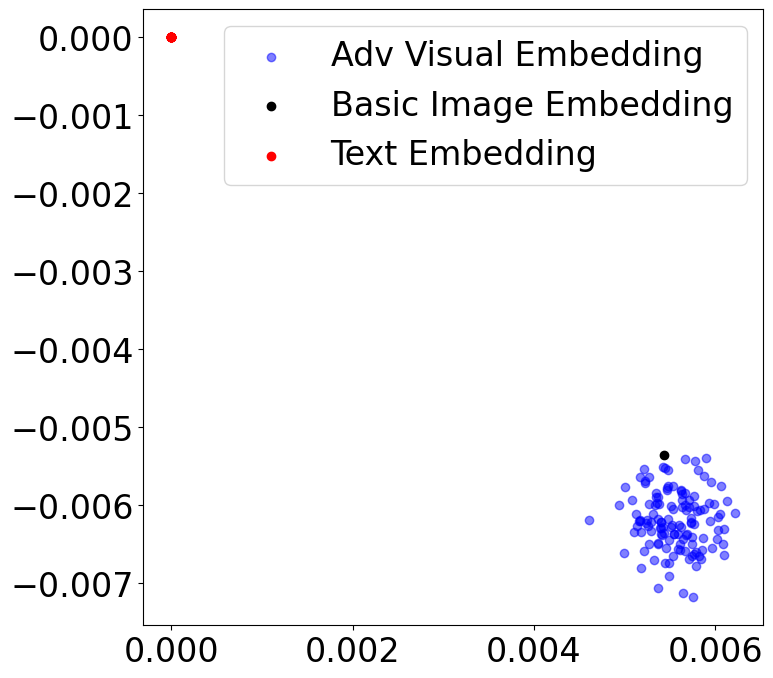}
}
\caption{Distributions of image-text pairs in CLIP and transform module embedding space.  }
\label{fig:embedding_space}
\vspace{-0.5cm}
\end{figure}

\begin{figure*}[t]
\centering
\subfigure[Basic image.]{
\label{Fig.sub.1}
\includegraphics[width=0.16\linewidth]{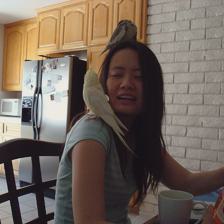}
}\vspace{-0.2cm}
\subfigure[Wikipedia.]{
\label{Fig.sub.2}
\includegraphics[width=0.16\linewidth]{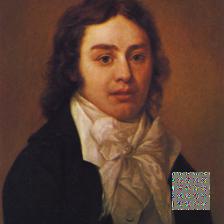}
}
\subfigure[Gaussian noise.]{
\label{Fig.sub.3}
\includegraphics[width=0.16\linewidth]{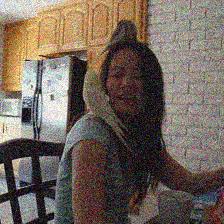}
}
\subfigure[Poisson  noise.]{
\label{Fig.sub.1}
\includegraphics[width=0.16\linewidth]{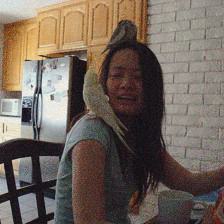}
}
\subfigure[Salt \& Pepper noise.]{
\label{Fig.sub.2}
\includegraphics[width=0.16\linewidth]{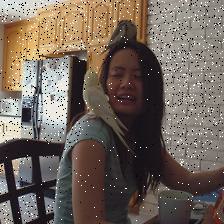}
}

\subfigure[Adversarial noise.]{
\label{Fig.sub.3}
\includegraphics[width=0.16\linewidth]{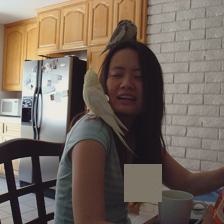}
}
\subfigure[Multiplicative noise.]{
\label{Fig.sub.1}
\includegraphics[width=0.16\linewidth]{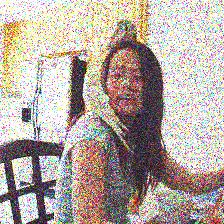}
}
\subfigure[Triangle patch.]{
\label{Fig.sub.2}
\includegraphics[width=0.16\linewidth]{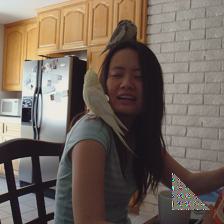}
}
\subfigure[Circle patch.]{
\label{Fig.sub.3}
\includegraphics[width=0.16\linewidth]{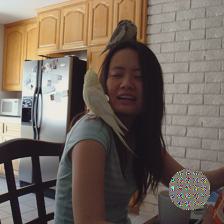}
}
\subfigure[Random Position.]{
\label{Fig.sub.3}
\includegraphics[width=0.16\linewidth]{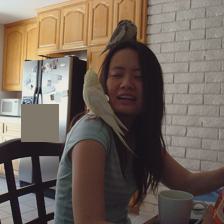}
}
\vspace{-0.5cm}
\caption{Visualization for different types of trigger generation strategies.}
\label{fig:side_by_side}
\end{figure*}


Figure \ref{fig:embedding_space} illustrated that the distance between adversarial triggers and basic text triggers in the transform embedding space was significantly reduced after implementing the transform module. 
The relative positions of adversarial triggers and basic image triggers became more concentrated, which evidenced that our transform module successfully narrowed the distance between adversarial triggers and basic text triggers, while increasing their similarity to basic image triggers. 
Consequently, the transform model could correctly alter the output results of adversarial triggers, aligning them with outputs of basic image triggers.

\subsection{Robustness}


We evaluated two adversarial attack scenarios to assess the robustness of AGATE, aligning with threat models in Section \ref{subsec:tm}.
 

\noindent\textbf{Scenario 1: Adversary with Partial Knowledge of Trigger Generation.}
The adversary was aware of the existence of triggers and attempted to fabricate false triggers to bypass the watermark verification, but was unaware of the specific trigger generation strategy.
We simulated the adversary by using various experimental groups in terms of types of trigger generation strategies (Figure \ref{fig:side_by_side}).
We set the strategy of adding adversarial, rectangular, and fixed-position noise patches to the basic trigger sampled from the Pascal dataset as a benchmark. 
Table \ref{tab:robust_trigger} showed that the adversary failed to bypass the watermark verification by creating forged triggers with different strategies, as reflected by $\textit{Res.} =0 $.

Specifically, we simulated the adversary lacked knowledge about which dataset the trigger originated from and what type of noise was added.
Similar performance was obtained when other types of noise were examined, e.g., GN, PN, SPN, and MN.
Results indicated a mismatch between the trigger and the text embedding, but 
failed to pass verification through the transform module. 
Results indicated a mismatch between the trigger and the text embedding, but failed to pass verification through the transform module. 
We analyzed that the transform module was trained on triggers generated by benchmark and only corrected the output results of these triggers.

In addition, we examined the performance when the adversary had knowledge of the trigger dataset and the type of noise added, but did not know the shapes and positions of adversarial patches.
Thus, adversaries might place different shapes at random positions on basic images. 
$\textit{Res}$\#1 $\textit{= True}$ indicated a large similarity between the trigger and the text embedding, evidencing the adversary's failure to forge a trigger that could bypass the watermark verification. 


\begin{table*}[!ht]
    \centering
    \caption{Comparison of classification task results across different trigger generation strategies in an adversarial scenario, where the adversary lacks complete knowledge of the trigger generation process.
    }
        \vspace{-0.2cm}
    \begin{tabular}{c|c|ccc|cccc|ccc}
    \hline
      \multicolumn{2}{c|}{Type} & $D(cos)$ & $D(euc)$  & Res\#1 & Module & $D(cos)$ & $D(euc)$ & Res\#2 & $\bigtriangleup_{cos}$  &  $\bigtriangleup_{euc}$  & Res.  \\ 
    \hline
    DataSets 
      & Wikipedia & 7.79 & 1.35 &  False   & $T_A$ & 14.06 & 1.30 &  False & +6.27 & -0.05 & 0 \\ 
      \hline
   \multirow{4}{*}{Noises} 
   & GN & 4.70 &  1.38 &  False   & $T_A$ & 37.89 & 1.10 &  False & +33.19 & -0.28 &  0\\ 
   & PN & 3.29 &  1.39 &  False & $T_A$ & 11.81 & 1.32 &  False & +8.52 & -0.07 &  0 \\
   & SPN & 9.90 &  1.34 &  False   & $T_A$ & 18.55 & 1.27 &  False & +8.65 & -0.07 &  0\\ 
   & MN & 5.02 &  1.37 &  False & $T_A$ & 13.50 & 1.30 &  False & +8.48 & -0.07 &  0 \\
    \hline
    
   \multirow{2}{*}{Shape} 
   & Triangle & 32.48 &  1.16 &  True   & $T_A$ & 76.40 & 0.67 &  True & +43.92 & -0.49 &  0\\ 
   & Circle & 28.97 &  1.19 &  True & $T_A$ & 77.08 & 0.66 &  True & +48.11 & -0.53 &  0 \\
    \hline

   Position  & Random & 34.25 &  1.14 &  True   & $T_A$ & 72.19 & 0.73 &  True & +37.94 & -0.41 &  0\\ 
 \hline
 \multicolumn{2}{c|}{Pascal, Adv, Rectangle, Fixed} 
    & 5.96 & 1.37 & False & $T_A$ & 88.63 & 0.47 &  True & +82.67 & -0.90 &  1\\ 
      \hline
    \end{tabular}
    \label{tab:robust_trigger}
\end{table*}


\begin{table*}[!t]
    \centering
    \caption{Comparison of classification task results across different triggers, models, and transform modules in an adversarial scenario, where the adversary lacks knowledge of the specific trigger and transform module details.
    }
    \vspace{-0.2cm}
    \begin{tabular}{c|cccc|cccc|ccc}
    \hline
        Trigger & Model & $D(cos)$ & $D(euc)$  & Res\#1 & Module & $D(cos)$ & $D(euc)$ & Res\#2 & $\bigtriangleup_{cos}$  &  $\bigtriangleup_{euc}$  & Res.  \\ 
    \hline 
   \multirow{5}{*}{Trigger A} 
      & $M_A$ & 3.11 & 1.39 &  False   & $T_A$ & 92.04 &  0.39 &  True & +88.93 & -1.00 & 1 \\ 
   & $M_A$ & 3.11 & 1.39 &  False & $T_B$ & 37.73 & 1.11 &  False & +34.62 & -0.28 &  0\\  
   & $M_A$ & 3.11 & 1.39 &  False & $T_C$ & 44.53 & 1.05 &  False & +41.42 & -0.34 &  0\\ 
   & $M_B$ & 34.09 & 1.14 &  True & $T_A$ & 55.95 & 0.93 &  True & +21.86 & -0.21 &  0\\ 
   & $M_C$ & 30.57 & 1.14 &  True & $T_A$ & 51.12 & 0.98 &  False & +20.55 & -0.19 &  1\\ 
      \hline
   Trigger B   & $M_A$ & 34.41 &  1.14 &  True   & $T_A$ & 48.62 & 1.00 &  False & +14.21 & -0.14 &  1\\ 
  Trigger C & $M_A$ & 34.33 &  1.14 &  True & $T_A$ & 47.92 & 1.00 &  False & +13.59 & -0.14 &  1 \\
      \hline
    \end{tabular}
    \label{tab:robust_module}
\end{table*}

\noindent\textbf{Scenario 2: Adversary Lacked Full Knowledge of Transform Module.}
The adversary remained unaware of the specific trigger and the detailed information about the transform module. 
To further evaluate the robustness of our framework, we considered enhancing the adversary's ability. 
In this case, the adversary had full knowledge of trigger generation.
The experiment results in Table \ref{tab:robust_module} demonstrated that despite the adversary's knowledge of the trigger generation and the transform module, they were still unable to successfully bypass the copyright verification, which highlighted the robustness of our framework against forgery attacks.
First, we observed distinct changes in $D(cos)$ and $D(euc)$ when Trigger $A$ was input into different CLIP models. 
However, $\textit{Res}$\#1 varied when different models were used.
Trigger $A$ output normally on $M_B$ and $M_C$ ($\textit{Res}$\#1 $=\textit{True}$), which meant it was not a suitable backdoor trigger for these models.
Thus, the adversary failed to directly apply model-specific triggers to other models.
Second, we observed that $\textit{Res}$\#1 and $\textit{Res}$\#2 remained consistent when we used different transform modules $T_B$ and $T_C$ after $M_A$. 
Therefore, the adversary was unable to forge a transform module that could bypass the copyright verification, which meant the transform module played a critical role in the verification process.
Combining these two aspects, attacks conducted by adversaries are ineffective whether they forge triggers or transform modules.




Two scenarios demonstrated that AGATE provided great robustness in preventing trigger forgery and in maintaining copyright verification against adversarial attacks.
Regardless of whether the adversary was aware of the trigger's generation method, the adversary could not successfully evade watermark verification. 
Model-specific triggers and model-related transform modules provided a stronger guarantee for our two-phase watermark verification.


\begin{figure}[t]
\centering
\subfigure[$D(cos)$]{
\label{Fig.sub.1_trigger}
\includegraphics[width=0.4\linewidth]{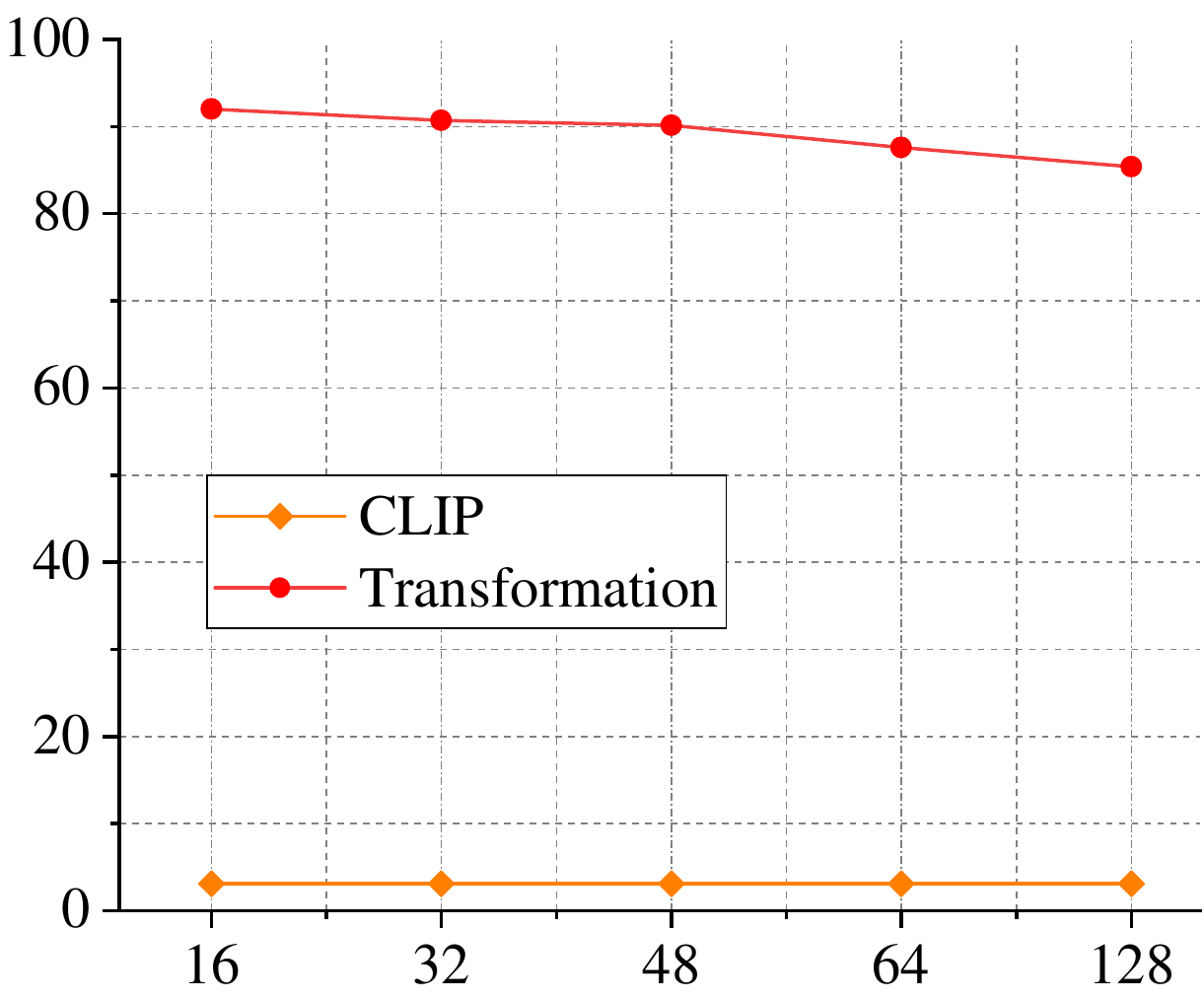}
}
\subfigure[$D(euc)$]{
\label{Fig.sub.2_trigger}
\includegraphics[width=0.4\linewidth]{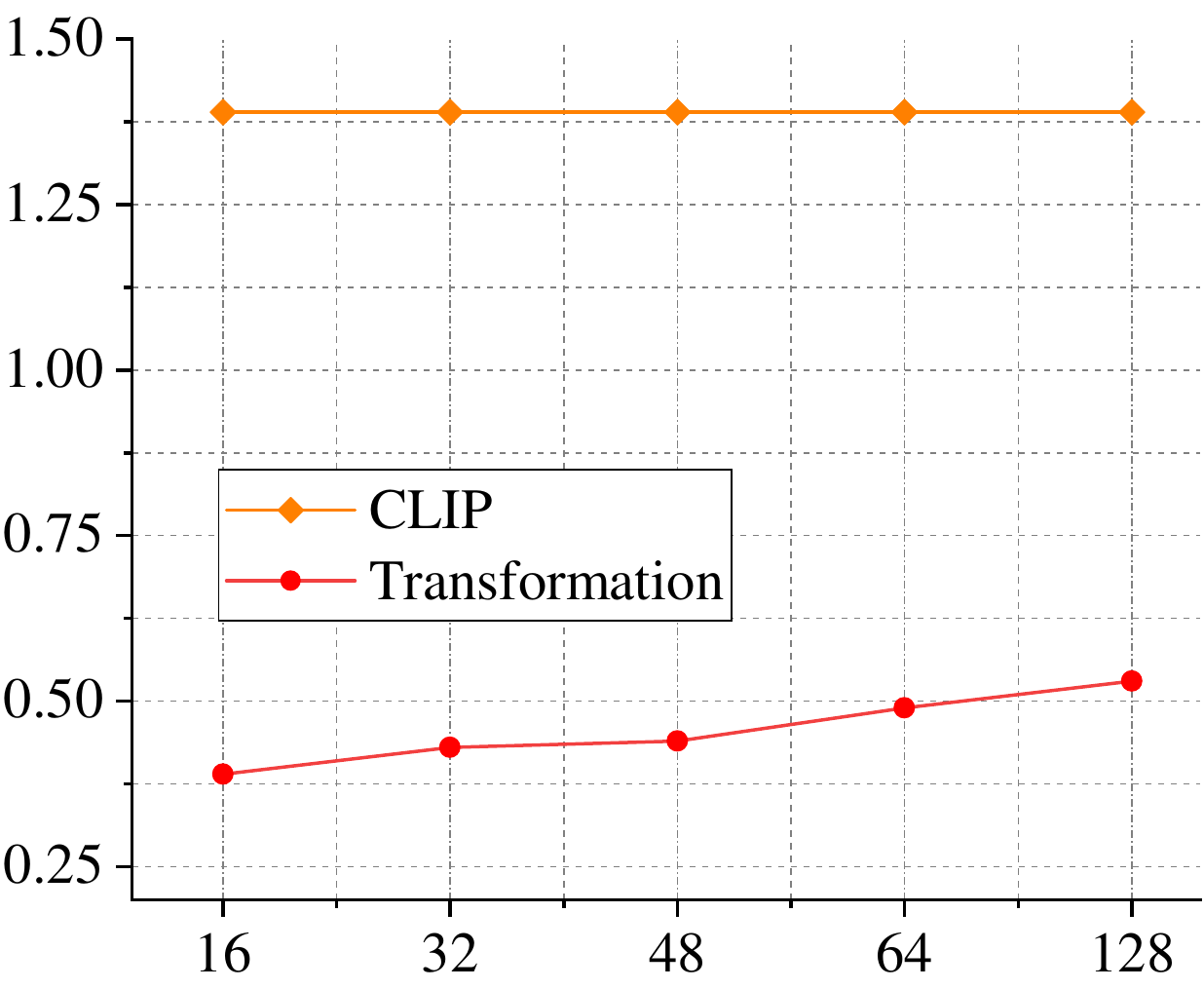}
}
    \vspace{-0.3cm}
\caption{Performance under different trigger numbers.}

\label{fig:trigger_number}

\end{figure}

\subsection{Impact of Trigger Number}


We investigated the impact of the number of triggers on the effectiveness of the transform module. 
Distance-based metrics were employed, including $D(cos) (\times 10^{-2})$  and $D(euc)$.
The evaluation focused on the difference in embedding distance after the transform module was applied, where a higher $D(cos)$ and lower $D(euc)$ value indicated a stronger effectiveness of transform module.

The results, presented in Figure \ref{fig:trigger_number}, demonstrated a clear trend. 
Increasing the number of triggers consistently reduced the effectiveness of the transform module. 
For example, $D(cos)$ reached 92.04 with 16 triggers, showing a significant increase of +88.93 compared to 3.11 before connecting the module. 
This indicated that 
were forcibly bound by the transform module. 
As the number of triggers increased, $D(euc)$ continued to rise, stabilizing at 0.53 when 128 triggers were used. 
The increase in $D(euc)$ implied that reducing the number of triggers helped ensure that trigger embeddings stayed close. 
Results showed that a lower trigger number strengthened the transform module’s ability to modify the embeddings.


Overall, the findings indicated that reducing the number of triggers enhanced the effectiveness of the transform module. 
However, reducing the number of triggers made it easier for adversaries to find specific triggers, greatly reducing the security of model copyright protection. 
This indicated that there existed a trade-off between maintaining high security and achieving maximum effectiveness according to the special requirements of application scenarios. 

\vspace{-0.3cm}
\section{Conclusion}
\label{sec:conclusion}

In this work, we addressed critical issues about copyright protection in multimodal AI models by proposing AGATE, a novel black-box backdoor watermarking framework. 
AGATE simplified the process of trigger selection by generating random adversarial noise, 
enhancing trigger security and stealth. 
Proposed transform module ensured accurate copyright verification by correcting outputs against adversarial attacks.  
Our work demonstrated that AGATE could efficiently protect copyrights across various multimodal models, offering an economical and effective solution. 





\bibliographystyle{ACM-Reference-Format}
\bibliography{sample-base}


\end{document}